\documentclass[twocolumn,aps, pra, nofootinbib,superscriptaddress,showkeys,showpacs,longbibliography]{revtex4-2}
\usepackage{amsmath,amsfonts,amssymb,amstext,amscd,amsthm,dsfont}
\usepackage{physics}
\usepackage{color}
\usepackage[colorlinks, citecolor=blue]{hyperref}
\usepackage{amsfonts}
\usepackage[font=small,format=plain,labelsep=period,labelfont=bf,textfont=normal,singlelinecheck=false,justification= raggedright]{caption}
\usepackage{braket}
\usepackage{graphicx}
\usepackage{subcaption}
\usepackage{times}
\usepackage{color}
\usepackage{soul,xcolor}
\usepackage{bm}
\usepackage[normalem]{ulem}
\usepackage{ragged2e}

\newcommand{\JIIT}{Jaypee Institute of Information Technology, A-10, Sector-62,Noida, UP-201309, India}
\newcommand{\CDOT}{Centre for Development of Telematics, Mehrauli, New Delhi - 110030, India}

\begin{document}

\title{Robust phase sensitivity in Mach-Zehnder interferometer using photon added and subtracted squeezed coherent state}
%===================================================

%===================================================
\author{Shivani Singh}
\email{shivani.singh@mail.jiit.ac.in}
\affiliation{%
\JIIT} 
\author{Priya Malpani}
\email{priya.ims07@gmail.com}
%\email{}

\affiliation{\JIIT}
\affiliation{\CDOT}
\author{Anirban Pathak}
\email{anirban.pathak@jiit.mail.ac.in}
\affiliation{\JIIT}
%===================================================================

%=========================================================================================
\begin{abstract} 

For the precision-based measurements, Mach-Zehnder interferometry is a widely used technique. There are various ways to enhance the precision of Mach-Zehnder interferometer (MZI), e.g., having a non-classical input state is one of the ways to enhance the precision of the phase estimation performed by MZI. The phase estimation performed by MZI is investigated here by considering that the input states of MZI are different combinations of photon added and subtracted squeezed coherent states (PASCS and PSSCS). Using quantum Fisher information, it is shown that the use of PASCS in both the input modes of MZI, provides the most precise estimate of the unknown phase. This system is also analyzed in two different measurement scenarios- single intensity detection (SID) and intensity difference detection (IDD). Systematic analysis has established that the intensity measurement might not be an optimal measurement scheme for phase estimation in MZI as phase and intensity correspond to non-commuting observables. The impact of the photon loss on the MZI-based phase estimation setup is also studied and it is found that PASCS is robust against photon loss, when the loss in MZI is low.

\end{abstract} 
%Keywords: Non-Gaussian states, Quantum Metrology, quantum Fisher information.

\maketitle
\vskip -0.7in
\noindent
%===================================================================

%===================================================================

\section{Introduction}

Estimation and sensitivity are the fundamental aspects of quantum metrology\,\cite{G11,G06}. In fact, the laws of quantum mechanics put fundamental limits on the precision of parameter estimation and sensitivity\,\cite{M63}. Keeping the relevance of precise measurement and sensitivity in general in mind, in the recent past, optical interferometry has been widely explored and studied for the focused investigation on the precise measurement of unknown phase using classical and non-classical (quantum) resources\,\cite{C81,B92,A19}. Specifically, non-classical resources are found to yield accuracy beyond classical capability\,\cite{M06}. With classical resources, the sensitivity of Mach-Zehnder interferometer (MZI) can go up to the standard
quantum limit (SQL), ($\triangle\phi_{SQL}\sim\frac{1}{\sqrt{\left\langle N\right\rangle }}$,
where $\left\langle N\right\rangle $ is the average number of photons). However, SQL can be breached to Heisenberg limit (HL), $\triangle\phi_{HL}\sim\frac{1}{\left\langle N\right\rangle }$ by using appropriate detection schemes and non-classical resources (e.g., NOON states \cite{israel2014supersensitive,W17},
squeezed state \cite{pezze2008mach}, cat state\,\cite{SM21}, two mode Fock state\,\cite{C03}). The biggest challenge of performing precise measurement with the non-classical states is that many of these states are very hard to generate and are susceptible to environmental noise and photon losses\,\cite{H25}. However, certain non-classical states can be generated and maintained relatively easily and one such state is squeezed state which can be produced by different types of matter field interaction processes including parametric down conversion (PDC) process. Interestingly, in MZI, use of a squeezed state can significantly enhance the phase sensitivity that reaches HL\,\cite{aasi2013enhanced,lang2013optimal}.

Gaussian states, such as squeeze state and squeezed displaced state, are of extreme importance in quantum optics and have been explored in extreme detail for phase sensitivity\,\cite{OL19,NL18,PM25}. Gaussian states are easy to engineer in optics in comparison to non-Gaussian states, but the  non-Gaussian states provide the optimal measurement needed for parameter estimation\,\cite{MP07}.  
Non-Gaussian states are also of extreme importance in quantum optics and computation for various other tasks\,\cite{AK25,KT06,SM86,HJ19}. 
In the recent past, the role of non-Gaussianity inducing operators have been proven to be an effective tool in improving the performance of Gaussian states. In quantum metrology, as well, non-Gaussian states play a very important role.  
After successful generation of photon added coherent state by Zavatta et al.  \cite{ZV04,ZV05}, a considerable amount of work has been done on non-Gaussian states e.g., superposition of coherent state with single-photon-added coherent state has been used to enhance the phase sensitivity in MZI\,\cite{SKS25}, phase sensitivity in lossy MZI via photon addition operation\,\cite{ZK26}, optimization of the phase sensitivity through photon-subtraction\,\cite{ZKZ26}, \textcolor{black}{or improving the phase sensitivity of SU(1, 1)-interferometer with photon-added squeezed vacuum light\,\cite{GY18}. Most of the existing works have either investigated photon added squeezed (coherent) states  or, photon subtracted squeezed (coherent) states. } In this article, we have investigated the role of non-Gaussian states, created by applying photon creation, and annihilation operations on squeezed coherent states called photon added squeezed coherent state (PASCS) and photon subtracted squeezed coherent state (PSSCS), respectively, in the improvement of phase sensitivity of MZI and compared them. We have observed that phase sensitivity not only depends on the input state, but it also depends on the detection scheme. Using quantum resources and appropriate detection scheme(s), e.g., intensity difference detection\,\cite{LS24}, single intensity detection \cite{ataman2018phase}, homodyne detection\,\cite{P06,gard2017nearly}, and parity detection \cite{seshadreesan2013phase}, one can reach up to HL. In what follows, we have compared the lower bound on phase sensitivity obtained using the quantum Cramer-Rao bound to the phase sensitivity obtained using two different detection schemes, intensity difference detection (IDD) and single intensity detection (SID), respectively. This article briefly elaborates on the basics of MZI and the engineered non-Gaussian states for which we have analyzed the phase sensitivity independent of measurement scheme using quantum Fisher information, which gives the maximum information about the unknown phase. Our analysis shows that the phase sensitivity of PASCS in both the input ports of MZI is maximum and the sensitivity of PSSCS in both the ports of MZI is minimum. For the combined input state of PASCS and PSSCS, the phase sensitivity is found to oscillate between the sensitivity of the maximum and minimum values.  
The phase sensitivity independent of measurement scheme obtained using quantum Fisher information is also the lower bound on the phase sensitivity and it is compared to the phase sensitivity obtained under IDD and SID schemes, respectively.
We have also analyzed the dependence of phase sensitivity for PASCS on multiple input parameters and have established the robustness of the PASCS state in both the input ports in the presence of photon loss.

The rest of the paper is organized as follows: Section \ref{sec:Model} introduces the standard MZI and the engineered non-Gaussian input states relevant to the present work. Section \ref{sec:phase_sensitivity} is dedicated to the study of phase sensitivity and the quantum Cramer-Rao bound. Subsequently, the results obtained here are being analyzed in Section \ref{sec:Observations} in an ideal scenario, and in Section \ref{sec:Loss} the impact of photon loss on the phase sensitivity is investigated. Finally, the work is concluded in Section \ref{sec:Conclusion}.

\section{Model}\label{sec:Model}

\subsection{Mach-Zehnder interferometer}
 
The standard Mach-Zehnder interferometer (MZI) consists of two beam splitters (BSs), two mirrors, and linear phase shifter(s) representing phase shift(s) in both modes of the interferometer, respectively. Given input states, which in our case are different combinations of PASCS and PSSCS, evolve using a $(50:50)$-BS followed by the combination of phase shifting operation and a second $(50:50)$-BS. Given, the photon annihilation operator $a_{1(2)}$ and creation operators $a^{\dag}_{1(2)}$ of mode-$1(2)$, respectively, the unitary operator form of the BS is $B = e^{-i\pi(a_1^{\dag}a_2 + a_1a_2^{\dag})/4}$ and the phase shift operator is $U(\theta_1,\theta_2) = e^{i(\theta_1 a^{\dag}_1 a_1 + \theta_2 a^{\dag}_2 a_2)}$, respectively. 
The phase shift operator can be decomposed into two terms, one representing the sum of the phases $(\theta_s)$ between the two modes and the other representing the relative phase difference $(\theta_d)$ between the two modes\,\cite{LC13}, given by
\begin{align}\label{eq:phase}
U(\theta_s,\theta) = e^{iN_s \theta_s/2} e^{iN_d \theta/2},
\end{align}
where $\theta_s = (\theta_1 + \theta_2)$ is the sum of the phases, and $\theta= (\theta_1 - \theta_2)$ is the relative phases difference between the two modes. The total number operator for the two modes is $N_s = (a^{\dag}_{1}a_{1} + a^{\dag}_{2}a_2)$, and number-difference operator is $N_d = (a^{\dag}_{1}a_{1} - a^{\dag}_{2}a_2)$, respectively.
The output state $\ket{\psi_{out}}$ of a standard MZI is 
\begin{align}\label{Output_state}
\ket{\psi_{out}} = BU(\theta_s,\theta)B\ket{\psi_{in}}.
\end{align}
where  $\ket{\psi_{in}}$ is the input state.

\begin{widetext}

\begin{figure}
\centering
\includegraphics[width=0.9\textwidth]{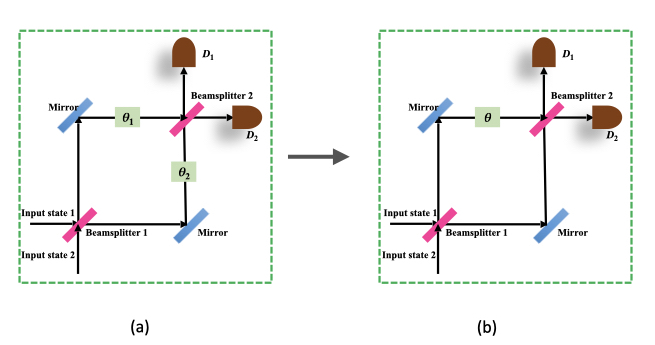}
\caption{(Color online) MZI setup consists of two mirrors, two $50:50$ beam splitters (Beamsplitter $1$ and Beamsplitter $2$) and two detectors ($D_1$ and $D_2$) where (a) shows the phases $\theta_1$ and $\theta_2$ in both the modes of MZI which is equivalent to 
(b) where $\theta$ is the relative phase difference between the two modes of the interferometer.}
\label{fig:Interferomery}
\end{figure}

\end{widetext}

In an interferometric experiment, we are only interested in the relative phase $\theta$ between the modes since, it is the only parameter which is relevant in phase estimation protocols as shown in Fig.\,\ref{fig:Interferomery}. When the relative phase is zero, the intensity (photon number) of the interferometer is maximum at the detector, due to constructive interference  between the modes.

The phase sensitivity in an interferometric setup depends upon the input state, the experimental setup, and the measurement strategy.
In the next section we have introduced the non-Gaussian states PASCS and PSSCS. Then we have used these states and its combination in MZI to establish the optimal input state for which the lowest bound on phase sensitivity is obtained.

\subsection{Photon added squeezed coherent state (PASCS) and photon subtracted squeezed coherent state (PSSCS)\protect\label{sec:state}}

The photon added squeezed coherent state\,\cite{thapliyal2017comparison}
is obtained by applying creation operator on squeezed coherent state.
The mathematical form of $m$ photon added squeezed coherent state
is given as
\begin{equation}
\left|\psi_{+}\right\rangle =N_{+}\left(\alpha,z,m\right)\hat{a}_{1}^{\dagger m}\left|\alpha,z\right\rangle ,\label{eq:PASCS}
\end{equation}
where $N_{+}\left(\alpha,z,m\right)$ is the normalization constant such that
\begin{align}
N_{+}\left(\alpha,z,m\right)&=\left[\left(\frac{\sinh2r}{4}\right)^{m}\sum_{l=0}^{m}\frac{\left(m!\right)^{2}\left(2\coth r\right)^{l}}{l!\left\{ \left(m-l\right)!\right\} ^{2}}\right. \nonumber \\
&\left.\times\left|H_{m-l}\left(A\right)\right|^{2}\right]^{-\frac{1}{2}},\label{eq:Norm_PASCS}
\end{align}
where
\[
A=i\frac{\alpha\exp\left(-\frac{i\phi}{2}\right)}{\sqrt{\sinh2r}}.
\]
The photon subtracted squeezed coherent state is obtained by applying
annihilation operator on squeezed coherent state. $m^{\prime}$ photon
subtracted squeezed coherent state is analytically given as
\begin{equation}
\left|\psi_{-}\right\rangle =N_{-}\left(\beta,z^{\prime},m^{\prime}\right)\hat{a}_{2}^{m^{\prime}}\left|\beta,z^{\prime}\right\rangle ,\label{eq:PSSCS}
\end{equation}
where $N_{-}\left(\beta,z^{\prime},m^{\prime}\right)$ is,
\begin{align}
N_{-}\left(\beta,z^{\prime},m^{\prime}\right)&=\left[\left(\frac{\sinh2r^{\prime}}{4}\right)^{m^{\prime}}\sum_{l=0}^{m^{\prime}}\frac{\left(m^{\prime}!\right)^{2}\left(2\tanh r^{\prime}\right)^{l}}{l!\left\{ \left(m^{\prime}-l\right)!\right\} ^{2}} \right. \nonumber \\
&\left. \times\left|H_{m^{\prime}-l}\left(B\right)\right|^{2}\right]^{-\frac{1}{2}},\label{eq:Norm_PSSCS}
\end{align}
where
\[
B=i\frac{\beta\exp\left(-\frac{i\phi^{\prime}}{2}\right)}{\sqrt{\sinh2r^{\prime}}}.
\]
\noindent
$\alpha$ and $\beta$ are displacement parameters, $z = (r\exp\left[-i\phi\right])$
and $z^{\prime} =(r^{\prime}\exp\left[-i\phi^{\prime}\right])$
are squeezing parameters. In the above and what follows, the subscripts $+$ and $-$ are used to indicate photon added and photon subtracted states, respectively.

In literature, various non-classical features of PASCS and PSSCS are
reported \cite{thapliyal2017comparison} such as higher-order antibunching,
higher-order sub Poissionian photon statistics, higher-order squeezing, and non-classical features revealed by
Klyshko's criterion. Analytic expressions for the higher-order moments necessary for the study of the presence of non-classical features revealed through the above-mentioned criterion and other moment-based criteria are given below for the quantum states of our interest:

\begin{align}
\left\langle \hat{a}_{1}^{\dagger p}\hat{a}_{1}^{q}\right\rangle _{+} &= N_{+}^{2}\left(\alpha,z,m\right)\exp\left[\frac{i\phi}{2}\left(p-q\right)\right]\nonumber \\
 & \times \sum_{j=0}^{{\rm min\left(p,q\right)}}\frac{\left(-1\right)^{j}p!q!\left(-i\right)^{p}i^{q}}{j!\left(p-j\right)!\left(q-j\right)!}\nonumber \\
 & \left(\frac{\sinh2r}{4}\right)^{m-j+\frac{p+q}{2}}\sum_{l=0}^{{\rm m-j+min\left(p,q\right)}}\frac{\left(2\coth r\right)^{l}}{l!} \nonumber \\
 & \frac{\left(m+p-j\right)!\left(m+q-j\right)!}{\left(m+p-j-l\right)!\left(m+q-j-l\right)!} \nonumber \\
 & H_{m+p-j-l}\left(A\right)H_{m+q-j-l}\left(A^{\star}\right).
\label{eq:HOM_PASCS}
\end{align}

\begin{equation}
\begin{array}{ccc}
\left\langle \hat{a}_{2}^{\dagger p}\hat{a}_{2}^{q}\right\rangle _{-} & = & N_{-}^{2}\left(\beta,z^{\prime},m^{\prime}\right)\exp\left[-\frac{i\phi^{\prime}}{2}\left(p-q\right)\right]\left(-i\right)^{p}i^{q}\\
 & \times & \sum_{l=0}^{{\rm m^{\prime}+min\left(p,q\right)}}\frac{\left(2\tanh r\right)^{l}}{l!}\frac{\left(m^{\prime}+p\right)!\left(m^{\prime}+q\right)!}{\left(m^{\prime}+p-l\right)!\left(m^{\prime}+q-l\right)!}\\
 & \times & \left(\frac{\sinh2r^{\prime}}{4}\right)^{m^{\prime}+\frac{p+q}{2}}H_{m^{\prime}+p-l}\left(B\right)H_{m^{\prime}+q-l}\left(B^{\star}\right).
\end{array}\label{eq:HOM_PSSCS}
\end{equation}

In the next section, we aim to study the phase sensitivity of MZI when the input states are various combinations of PASCS and PSSCS. For  the  analytical calculation of the phase sensitivity, higher order moments are used.

\section{Phase sensitivity\protect\label{sec:phase_sensitivity}}

In this section, we aim to systematically investigate phase sensitivity in MZI. 
The relative phase $\theta$ can be estimated by measuring the intensity of the two modes. Here, we may note that the detection schemes considered here are IDD and SID.  

During the detection process, the general form of the observable is given by, $\hat{G}\left(\theta\right)=P\hat{a}_{3}^{\dagger}\hat{a}_{3}+Q\hat{a}_{4}^{\dagger}\hat{a}_{4},$ 
where $a_3 (a_3^{\dagger})$ and $a_4(a_4^{\dagger})$ are the annihilation(creation) operators at output, obtained using Eq.\,\eqref{Output_state}. When $P=Q=-1$, we obtain intensity difference detection (IDD) and when $P=1,Q=0$ and $P=0,Q=1$,
we obtain single intensity detection (SID)\,\cite{SA18, SKS25}.
The expectation value of operator $\left\langle \hat{G}\left(\theta\right)\right\rangle $ is 
defined as, 
\begin{equation}
\left\langle \hat{G}\left(\theta\right)\right\rangle =P\left\langle \hat{a}_{3}^{\dagger}\hat{a}_{3}\right\rangle +Q\left\langle \hat{a}_{4}^{\dagger}\hat{a}_{4}\right\rangle .\label{op}
\end{equation}
Phase sensitivity for the given detection schemes, IDD and SID, in terms of operator $\left\langle \hat{G}\left(\theta\right)\right\rangle $ can be
given by\,\cite{SA18, SKS25},
\begin{equation}
\left(\triangle\theta\right)^{2}=\frac{\left(\triangle\hat{G}\left(\theta\right)\right)^{2}}{\left|\frac{\partial\left\langle \hat{G}\left(\theta\right)\right\rangle }{\partial\theta}\right|^{2}}\label{eq:phase}
\end{equation}

\begin{equation}
\left(\triangle\hat{G}\left(\theta\right)\right)^{2}=\left\langle \hat{G^{2}}\left(\theta\right)\right\rangle -\left\langle \hat{G}\left(\theta\right)\right\rangle ^{2}\label{eq:phase1}
\end{equation}
Here, $\left\langle \hat{G^{2}}\left(\theta\right)\right\rangle $
and $\left\langle \hat{G}\left(\theta\right)\right\rangle $ are the expectation values, such that 
\begin{equation}
\left\langle \hat{G}\left(\theta\right)\right\rangle =l_{1}E_{1}+l_{2}E_{2}-l_{3}E_{3}\label{avh},
\end{equation}
\begin{equation}
\left|\frac{\partial\left\langle \hat{G}\left(\theta\right)\right\rangle }{\partial\theta}\right|=\left|l_{3}\left(E_{1}-E_{2}\right)-l_{3}^{\prime}E_{3}\right|\label{avg}
\end{equation}
and
\begin{align}\label{eq:avg2}
\left\langle \hat{G^{2}}\left(\theta\right)\right\rangle  &= l_{1}^{2}E_{5}+l_{2}^{2}E_{4}+\left(l_{3}^{2}+l_{1}^{2}\right)E_{1}+\left(l_{3}^{2}+l_{2}^{2}\right)E_{2}  \nonumber \\
& +  l_{3}^{2}E_{6} + 2\left(l_{1}l_{2}+l_{3}^{2}\right)E_{7} - l_{3}\left(l_{1} +  l_{2}\right)E_{3} \nonumber \\
& - 2l_{2}l_{3}E_{8}-2l_{3}l_{1}E_{9} 
\end{align}
where, 
\begin{align}
l_{1} &= P\sin^{2}\left(\frac{\theta}{2}\right)+Q\cos^{2}\left(\frac{\theta}{2}\right), \nonumber \\
l_{2} &= Q\sin^{2}\left(\frac{\theta}{2}\right)+P\cos^{2}\left(\frac{\theta}{2}\right), \nonumber \\
l_{3} &= \frac{1}{2}\left(P-Q\right)\sin\left(\theta\right), \nonumber \\
l_{3}^{\prime} &= \frac{1}{2}\left(P-Q\right)\cos\left(\theta\right), \nonumber
\end{align}

\noindent
and  $E_{1} = \left\langle \hat{a}_{1}^{\dagger}\hat{a}_{1}\right\rangle$, $E_{2} = \left\langle \hat{a}_{2}^{\dagger}\hat{a}_{2}\right\rangle$, $
E_{3} = \left\langle \hat{a}_{1}\hat{a}_{2}^{\dagger}\right\rangle$, $ E_{4} = \left\langle \hat{a}_{2}^{2\dagger}\hat{a}_{2}^{2}\right\rangle$, $
E_{5} = \left\langle \hat{a}_{1}^{2\dagger}\hat{a}_{1}^{2}\right\rangle $, $E_{6} = \left\langle \hat{a}_{1}^{2}\hat{a}_{2}^{2\dagger}\right\rangle +\left\langle \hat{a}_{1}^{2\dagger}\hat{a}_{2}^{2}\right\rangle$, $ E_{7} = \left\langle \hat{a}_{1}^{\dagger}\hat{a}_{1}\hat{a}_{2}^{\dagger}\hat{a}_{2}\right\rangle$, $E_{8} = \left\langle \hat{a}_{1}\hat{a}_{2}^{2\dagger}\hat{a}_{2}\right\rangle +\left\langle \hat{a}_{1}^{\dagger}\hat{a}_{2}^{\dagger}\hat{a}_{2}^{2}\right\rangle$, $ E_{9} = \left\langle \hat{a}_{1}^{\dagger}\hat{a}_{1}^{2}\hat{a}_{2}^{\dagger}\right\rangle +\left\langle \hat{a}_{1}^{2\dagger}\hat{a}_{1}\hat{a}_{2}\right\rangle$.

The lower bound on the phase sensitivity, independent of the detection schemes, is given by Quantum Cramer-Rao bound (QCRB), which states that the phase sensitivity in a system is lower bounded by quantum Fisher information (QFI) $I_{dd}$ in the system, given by
\begin{equation}
\triangle\theta_{QCRB}=\frac{1}{\sqrt{I_{dd}}}. \label{eq:QCRB}
\end{equation}
Significant amounts of work have been done on QFI\,\cite{MR12,M09,A20}. 
If the output, before detection, is a pure state then QFI for single parameter estimation is,
\begin{equation}
I_{dd}=4\left(\left\langle \partial_{\theta}\psi\shortmid\partial_{\theta}\psi\right\rangle -\left\langle \partial_{\theta}\psi\shortmid\psi\right\rangle \left\langle \psi\shortmid\partial_{\theta}\psi\right\rangle \right)\label{eq:QCRB2}.
\end{equation}
Therefore, the QFI for the estimation of the relative phase $\theta$ is
\begin{align}\label{eq:QFI_d}
I_{dd}=2E_{7}+E_{1}+E_{2}-E_{6}+E_{10}^{2}
\end{align}
such that the quantity $E_{10}=\left\langle \hat{a}_{1}^{\dagger}\hat{a}_{2}\right\rangle -\left\langle \hat{a}_{1}\hat{a}_{2}^{\dagger}\right\rangle $. The phase sensitivity obtained using QFI, Eq.\,\eqref{eq:QFI_d}, is independent of the parameter to be estimated i.e., relative phase $\theta$. It lower bounds the precision obtained in Eq.\,\eqref{eq:phase}, using the observable $G(\theta)$, for the estimation of phase $\theta$. 

\section{Observations \protect\label{sec:Observations}}

In this section, we plan to study the lower bound on the phase sensitivity, obtained using QFI, for different combinations of PASCS and PSSCS as input states. Specifically, we aim to identify the input state for which we obtain the lowest bound on the phase sensitivity and compare it with the phase sensitivity obtained when the detection scheme is IDD or SID. The lower bound in MZI is given by the quantum Carmer-Rao bound, described in Eq.\,\eqref{eq:QCRB}, which is inversely proportional to the square root of the QFI.  

\begin{figure}
\centering
\includegraphics[width=0.5\textwidth]{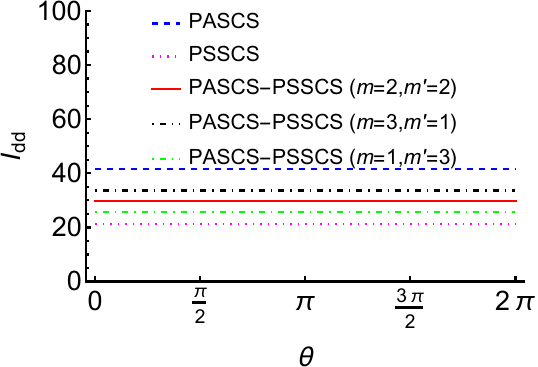}
\caption{(Color online) QFI $I_{dd}$ for the estimation of relative phase $\theta$ with respect to $\theta$. The number of photon added or subtracted in PASCS or PSSCS case are $m=m^{\prime}=2$, respectively. Here, $r=0.01$, $r^{\prime}=0.01$, $\phi=\pi$, $\phi^{\prime}=\pi$, and $\alpha=\beta=3$. }
\label{fig:QFI_compare}
\end{figure}

Our observation shows that the QFI $(I_{dd})$ is maximum when the input states are PASCS in both the modes and minimum when the input states are PSSCS in both the modes. $I_{dd}$ for a combined input state of PASCS-PSSCS (one mode has PASCS input state and other mode has PSSCS input state in MZI), oscillates between the maximum and minimum $I_{dd}$, respectively. It is illustrated in Fig.\,\ref{fig:QFI_compare}. 
We observe that QFI increases if the added photon (PASCS) is higher than subtracted photon (PSSCS) in case of PASCS-PSSCS input state. Another important observation is that  QFI is independent of the relative phase $\theta$, consistent with the analytical result obtained in Eq.\,\eqref{eq:QFI_d}, for all the combinations PASCS and PSSCS. 
The numerical analysis reveals that PASCS in both the input ports of MZI provides maximum phase sensitivity therefore the in the next section, the rest of the analysis will be done for the case of PASCS in both the input ports of the MZI. 

\subsection{QCRB for PASCS input state in MZI}\label{QCRB}

When the inputs of both the modes of MZI are PASCS, Eq.\,\eqref{eq:PASCS}, the combined input state is defined as the tensor product of two PASCS as
\begin{equation}
\left|\psi_{+}\right\rangle \otimes\left|\psi_{+}\right\rangle =\hat{a}_{1}^{\dagger m}\left|\alpha,z\right\rangle \otimes\hat{a}_{2}^{\dagger m^{\prime}}\left|\beta,z^{\prime}\right\rangle ,\label{BS-1}
\end{equation}

QCRB for this input state is given by Eq.\,\eqref{eq:QCRB}, where it provides the lowest bound on the phase sensitivity of relative phase $\theta$ compared to other combinations of PASCS and PSSCS, independent of the measurement scheme. PASCS has three independent parameters, one associated with the displacement term $\alpha$, and two associated with the squeezing term $z = r\exp(i\phi)$. Since we have PASCS state in each mode of MZI, the total number of independent parameters in the input states is six, including $\beta$ and $z^{\prime} = r^{\prime} \exp(i\phi^{\prime})$.

\begin{figure}
\begin{centering}
\begin{tabular}{cc}
\includegraphics[width=0.5\textwidth]{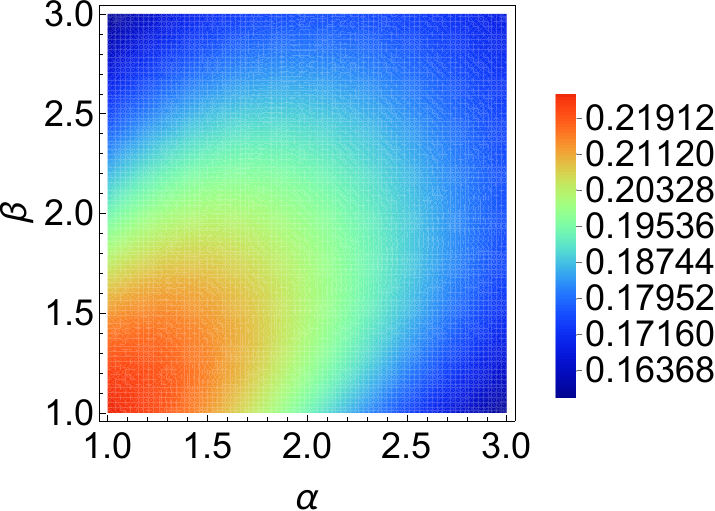} \tabularnewline 
\tabularnewline
(a) \\
\\
\includegraphics[width=0.5\textwidth]{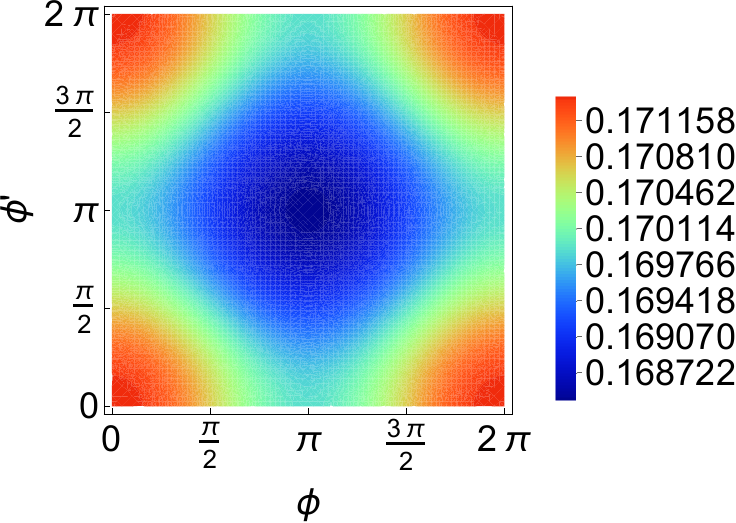} \tabularnewline
\tabularnewline
(b) \tabularnewline 
\\
\includegraphics[width=0.5\textwidth]{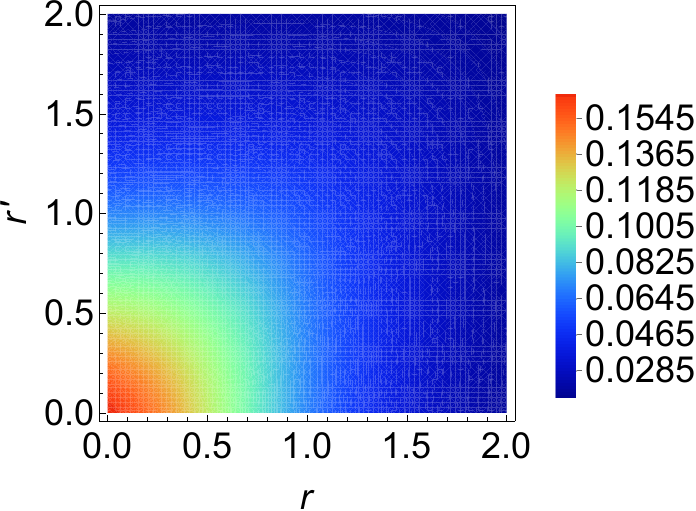} \tabularnewline
\tabularnewline
(c) \tabularnewline
\end{tabular}
\par\end{centering}
\caption{\protect\label{fig:QFI_PASCS-PASCS}(Color online) Phase sensitivity using QCRB $\Delta\theta_{QCRB}$ with respect to (a) displacement parameters $\alpha$ and $\beta$, (b) the phase associated with the squeezing parameters $\phi$ and $\phi^{\prime}$, and (c) squeezing amplitude $r$ and $r^{\prime}$ when PASCS is the input state on both the ports when $m=2$ and $m^{\prime}=2$.  }
\end{figure} 

Fig.\,\ref{fig:QFI_PASCS-PASCS}-(a) shows the phase sensitivity using QCRB with respect to the displacement parameter associated with coherent terms $\alpha$ and $\beta$. It shows that the phase sensitivity increases with the parameters $\alpha$ and $\beta$, implying that while performing the phase estimation protocols,  having a coherent term leads to better estimation of the unknown phase. However, the phase resolution protocol performs best when the quantumness in the protocol increases with the squeezing term, which has two parameters associated with it, $r$ and $\phi$, such that the squeezing factor is $z=re^{i\phi}$. Fig.\,\ref{fig:QFI_PASCS-PASCS}-(b) and (c) illustrate the phase sensitivity using QCRB $\Delta \theta_{QCRB}$ with respect to $\phi$ and $\phi^{\prime}$, and $r$ and $r^{\prime}$, respectively, where $z=re^{i\phi}$ and $z^{\prime} = r^{\prime}e^{i\phi^{\prime}}$ are the squeezing factors in each mode of the MZI, respectively. Numerical analysis reveals that the phase sensitivity is lowest when $\phi$, in both the modes, is an integral multiple of $\pi$ while in the case of $r$, it approaches the lowest value as the value of $r$ increases. 

QCRB provides the maximum achievable precision (lowest possible phase sensitivity)  when estimating unknown phase, independent of the measurement techniques\,\cite{M09}, and our analysis suggests that the precision in the MZI  increases with the squeezing and displacement parameters in the input state while it is independent of the unknown relative phase $\theta$. In the next section, we will compare the results of the QCRB with the results obtained using two of the well-known detection schemes, IDD and SID. Through our analysis, we will establish the conditions where one can reach the ultimate bound (obtain using QCRB) using the intensity measures, IDD and SID.

\subsection{Intensity measurement schemes}\label{IMS}

\begin{figure}
\begin{centering}
\begin{tabular}{cc}
\includegraphics[width=0.5\textwidth]{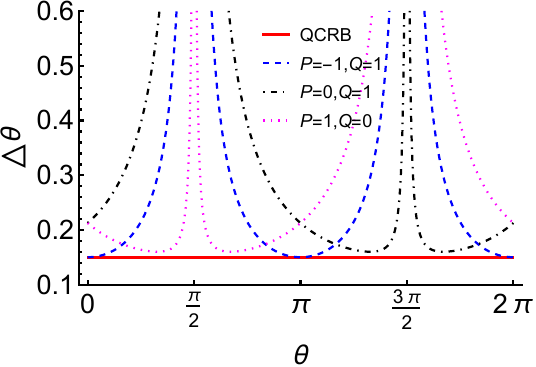} \tabularnewline 
\tabularnewline
(a) \\
\\
\includegraphics[width=0.5\textwidth]{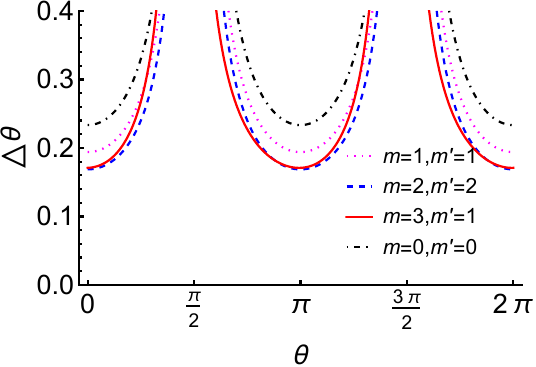} \tabularnewline
\tabularnewline
(b) \tabularnewline
\end{tabular}
\par\end{centering}
\caption{\protect\label{fig:PASCS-PASCS}(Color online) Phase sensitivity $\Delta \theta$ with respect to $\theta$. PASCS is the input state on both the ports (a) for different detection schemes: SID (when $P=0(1)$ and $Q=1(0)$) and IDD (when $P=-1$ and $Q=1$) when $m=2,m^{\prime}=2$ and red line represents $\Delta \theta_{QCRB}$, calculated using QFI, (b) for different combinations of the number of added photons $(m,m^{\prime})$ when detection scheme is IDD ($P=-1$ and $Q=1$). Here, $r=0.01,r^{\prime}=0.01$, $\phi=\pi$ and $\phi^{\prime}=\pi$ and $\alpha=\beta=3$.}
\end{figure}

Intensity measure is one of the easily achieved detection techniques. In this section, we will be focusing on SID and IDD detection schemes when the input state of MZI is PASCS, given by Eq.\,\eqref{BS-1}. Fig.\,\ref{fig:PASCS-PASCS}-(a) compares phase sensitivity in IDD, where $P=-1$  and $Q=1$, and SID where $P=0(1)$  and $Q=1(0)$, with respect to the unknown relative phase $\theta$, when the rest of the input parameters are arbitrary such as squeezing phase $\phi = \pi$ and squeezing amplitude $r = 0.01$ and coherent parameter is $3$ in both the modes. \textcolor{black}{IDD reaches the lowest bound on the phase sensitivity i.e., $\Delta \theta_{QCRB}$, when the relative phase $\theta$ is an integral multiple of $\pi$ and blows up when the relative phase $\theta$ reaches close to an odd multiple of $\pi/2$. In SID scheme, phase sensitivity blows up at an odd multiple of $\pi/2$ as well, but reaches minimum phase sensitivity for $P=1,Q=0$ when the relative phase is $(n+1/2)\pi < \theta < (n+1)\pi$ where $n=\{1,3,5,...\}$, and for $P=0,Q=1$ when the relative phase is $(n'+1/2)\pi<\theta< n'\pi$ where $n'=\{0,2,4,...\}$, respectively.}

Fig.\,\ref{fig:PASCS-PASCS}-(b) shows phase sensitivity $\Delta \theta$ in IDD with respect to  unknown phase $\theta$ for different combination of the number of photons added in the each mode of MZI. It shows that when the total number of  added photons i.e., $(m+m^{\prime})$ in the input state increases $\Delta \theta$ for the IDD decreases. Keeping $(m+m^{\prime})$ constant in the MZI, gives the same minimum phase sensitivity $\Delta \theta$ implying photon addition process actually improves the phase sensitivity and only the lowest value of the phase sensitivity in IDD is affected by the the combined number of added photons $(m+m^{\prime})$. 

\begin{figure}
\begin{centering}
\begin{tabular}{cc}
\includegraphics[width=0.4\textwidth]{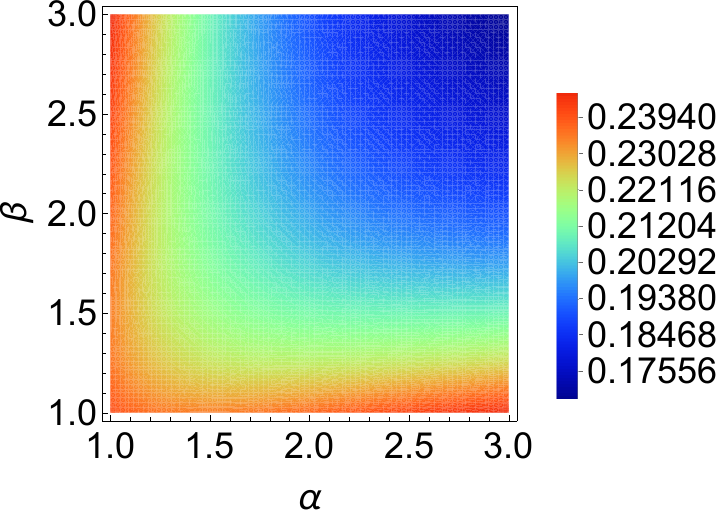} \tabularnewline 
\tabularnewline
(a) \\
\\
\includegraphics[width=0.4\textwidth]{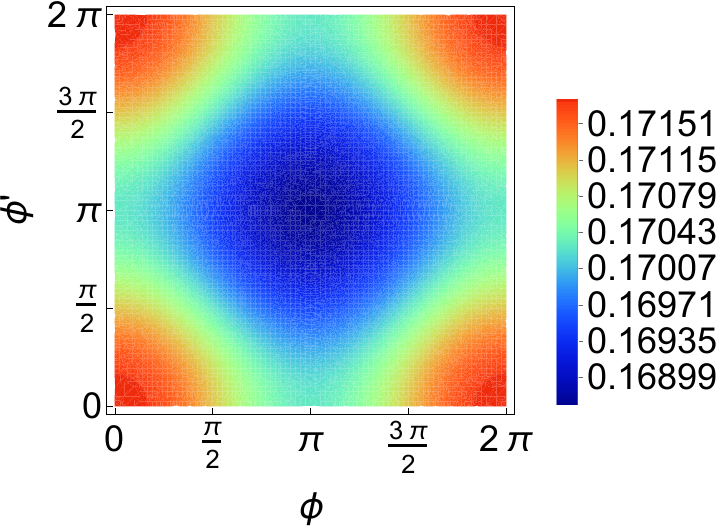} \tabularnewline
\tabularnewline
(b) \tabularnewline 
\\
\includegraphics[width=0.4\textwidth]{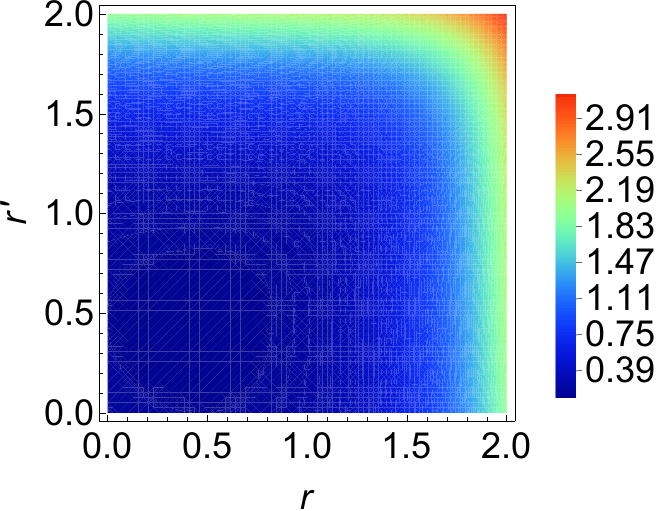} \tabularnewline
\tabularnewline
(c) \tabularnewline
\end{tabular}
\par\end{centering}
\caption{\protect\label{fig:IDD_PASCS-PASCS} (Color Online) Phase sensitivity $\Delta\theta$ when the measurement scheme is IDD where $P=1$ and $Q=-1$ with respect to (a) displacement parameters $\alpha$ and $\beta$, (b) the phase associated with the squeezing parameters $\phi$ and $\phi^{\prime}$, and (c) squeezing amplitude $r$ and $r^{\prime}$ when PASCS is the input state on both the ports when $m=2$ and $m^{\prime}=2$. }
\end{figure} 

Fig.\,\ref{fig:IDD_PASCS-PASCS} shows the contour plot for the phase sensitivity with respect to various input parameter when the detection scheme is IDD. Fig.\,\ref{fig:IDD_PASCS-PASCS}-(a) illustrates $\Delta \theta$ with respect to displacement parameters $\alpha$ and $\beta$, and it shows that, increasing the displacement parameters in both modes enhance the phase precision (reduces the phase sensitivity) implying that the higher the displacement parameter, better is the phase sensitivity, which is same as our observation for QCRB with respect to displacement parameters in the previous section except in IDD scheme $\Delta \theta$ decreases when both the displacement parameters $\alpha$ and $\beta$ are large while in QCRB, $\Delta \theta$ is reduced when either of the displacement parameters $\alpha$ or $\beta$ are large. 

Fig. \ref{fig:IDD_PASCS-PASCS}-(b) shows the phase sensitivity with respect to the squeezing phase $\phi$ and $\phi^{\prime}$ in each mode, respectively. The behavior of phase sensitivity with respect to squeezed phase is the same as QCRB.  i.e., $\Delta \theta$ is minimum when $\phi = \phi^{\prime}=\pi$. Fig. \ref{fig:IDD_PASCS-PASCS}-(c) shows phase sensitivity with respect to the squeezing amplitude $r$ which should be small for minimum phase sensitivity. This observation is in contrast to the observations made in the context of QCRB discussed in the previous section. This is due to the Heisenberg uncertainty relation between the phase and the intensity (number of photon) i.e., $\Delta\theta\Delta N \geq 1/2$\,\cite{PB89}. Thus, when the intensity measurement is precise, the error in phase measurement is high i.e., phase sensitivity in the system is high. The intensity in MZI increases exponentially with the squeezing parameter $r$ as $I \approx \sinh^2(r)$. Since we are measuring the intensity in the IDD scheme, the phase sensitivity in IDD is in contrast to QCRB. Phase sensitivity in IDD increases with the squeezing amplitude $r$. It further implies that the intensity measurement is easily achieved, but not the optimal measurement scheme to estimate the unknown phase in MZI, when the input is PASCS with a large squeezing parameter.

\begin{figure}
\begin{centering}
\begin{tabular}{cc}
\includegraphics[width=0.4\textwidth]{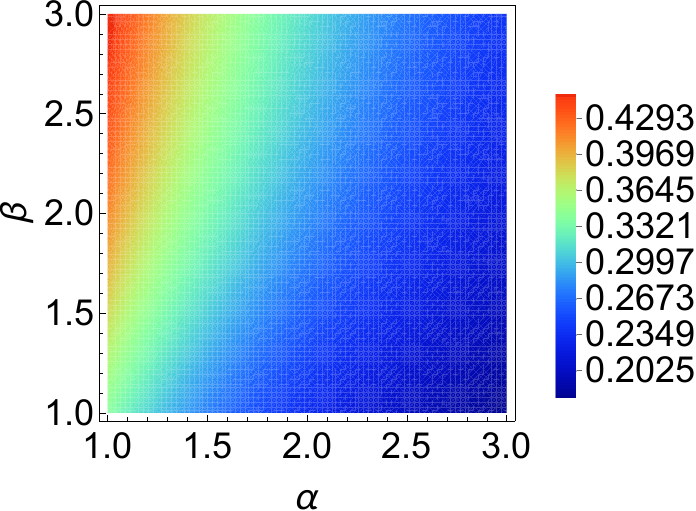} \tabularnewline 
\tabularnewline
(a) \\
\\
\includegraphics[width=0.4\textwidth]{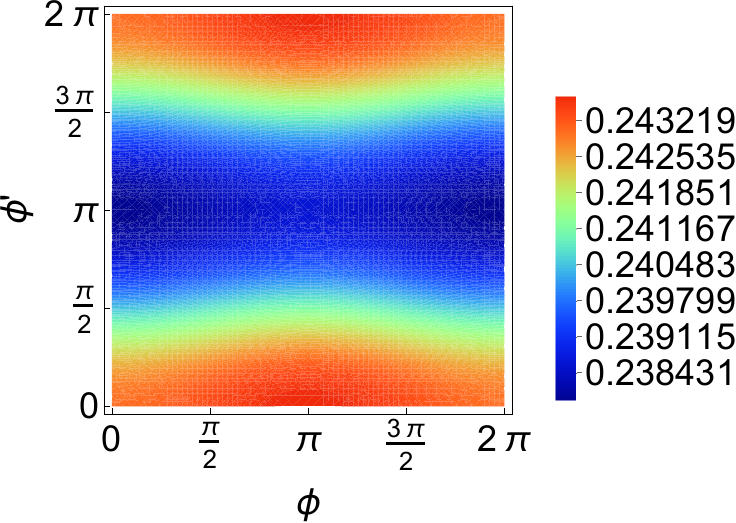} \tabularnewline
\tabularnewline
(b) \tabularnewline 
\\
\includegraphics[width=0.4\textwidth]{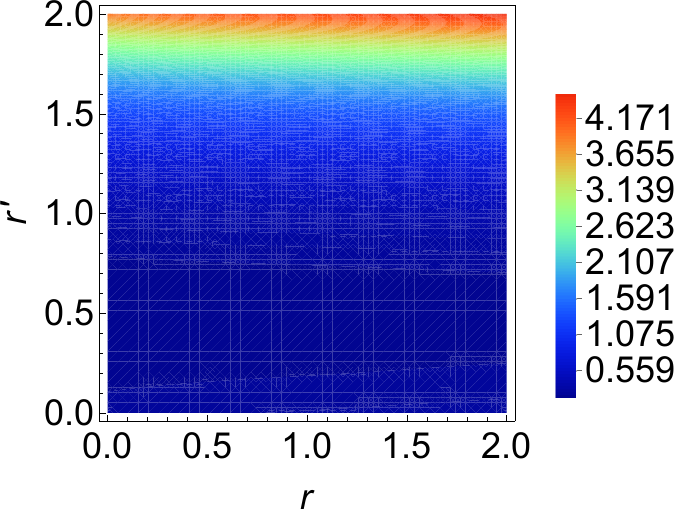} \tabularnewline
\tabularnewline
(c) \tabularnewline
\end{tabular}
\par\end{centering}
\caption{\protect\label{fig:SID_PASCS-PASCS} (Color Online) Phase sensitivity $\Delta\theta$ when the measurement scheme is SID where $P=0$ and $Q=1$ with respect to (a) coherent parameters $\alpha$ and $\beta$, (b) the phase associated with the squeezing parameters $\phi$ and $\phi^{\prime}$, and (c) squeezing amplitude $r$ and $r^{\prime}$ when PASCS is the input state on both the ports when $m=2$ and $m^{\prime}=2$. }
\end{figure} 

We establish this observation with another type of intensity detection scheme called SID. Fig.\,\ref{fig:SID_PASCS-PASCS} shows the phase sensitivity in SID when the detector is at the second port of MZI (implying $P=0$ and $Q=1$) with respect to input parameters. Fig.\,\ref{fig:SID_PASCS-PASCS}-(a) shows that the $\Delta \theta$ in SID reduces with the displacement parameter in the undetected mode while in the detected mode it only reduces when the displacement parameter is large and equal to the  parameter in the undetected mode. 
Fig.\,\ref{fig:SID_PASCS-PASCS}-(b) shows the phase sensitivity with respect to squeezed phase $\phi$. The squeezed phase of undetected mode does-not affect the $\Delta \theta$ while the squeezed phase of detected mode performs better when $\phi$ is near the value $\pi$. Fig.\,\ref{fig:SID_PASCS-PASCS}-(c) shows the phase sensitivity with respect to squeezing amplitude $r$. Here also  we find that the undetected mode does not affect the $\Delta \theta$, but in the detected mode, $\Delta \theta$ increases with $r$, in a manner similar to what is observed for IDD, but phase sensitivity in SID is higher than that in IDD due to information loss in the undetected mode.

\section{Phase sensitivity in presence of photon-loss}\label{sec:Loss}

\begin{figure}
\centering
\includegraphics[width=0.5\textwidth]{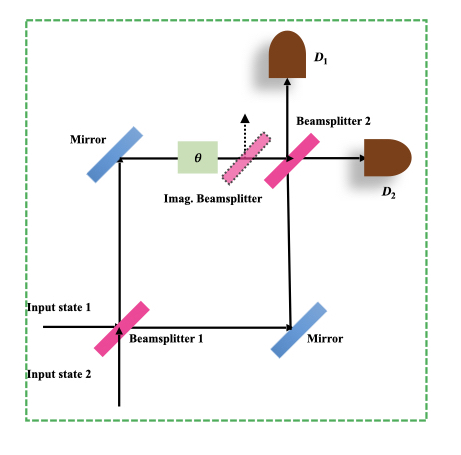}
\caption{ (Color Online) MZI setup with photon loss in mode-1, represented by an imaginary beam-splitter. Input state 1 and input state 2 are PASCS state, respectively and phase to be estimated is $\theta$ which is the relative phase difference between the arms of the interferometer. $D_1$ and $D_2$ are the detectors at both the modes, respectively.}
\label{fig:MZI_loss}
\end{figure}

In this section, we investigate phase sensitivity in the presence of photon loss. For simplicity, we have focused on the photon loss in only one mode of MZI as shown in Fig.\,\ref{fig:MZI_loss}. The photon loss is modeled using an imaginary beam-splitter (BS) with the transmittance $0 \geq t \leq 1$. Complete photon loss is represented by $t=0$, and zero loss corresponds to $t=1$.  The dynamics of MZI with photon loss is characterized through Kraus operator formalism\,\cite{EM11, CYZ22, ZKZ25}. The QFI in the presence of photon loss is given by,
\begin{align}
Idd_{L} = \frac{4tI_{dd}\langle n_a \rangle}{(1-t)I_{dd} + 4t\langle n_a \rangle},
\end{align}
where $t$ is the transmittance of the fictitious beam splitter, $I_{dd}$ is the QFI in the ideal case and $\langle n_a \rangle = \langle a_{1}^{\dag}a_1 \rangle$ is the internal average photon number of mode with photon loss $a_1$. Loss is maximum when the transmittance $t<<1$. 

\begin{figure}
\centering{}
\includegraphics[width=0.5\textwidth]{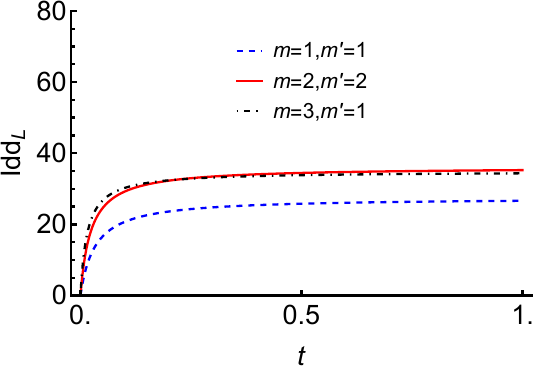}
\caption{\protect\label{fig:QFI_Loss}  (Color Online)  QFI to estimate the relative phase $\theta$ in presence of photon loss with respect to transmission coefficient $t$ in MZI, where $t=0$ corresponds to complete loss and $t=1$ corresponds zero photon-loss. The input state is PASCS with different combination of number of added photons in each mode $m$ and $m'$. Here, $r=0.01$, $r^{\prime}=0.01$, $\phi=\pi$, $\phi^{\prime}=\pi$, and $\alpha=\beta=3$. }
\end{figure}

Fig.\,\ref{fig:QFI_Loss} shows $Idd_{L}$ to estimate $\theta$ in presence of loss in one mode with respect to transmission coefficient $t$. It shows that QFI is robust against photon loss, when the photon loss is small i.e., when $t \rightarrow 1$. The phase sensitivity in presence of photon loss is given by QCRB Eq.\,\eqref{eq:QCRB},
\begin{align}
\Delta^2 \theta_{QCRB} &= \frac{1}{Idd_L} \nonumber \\
&= \frac{(1-t)}{4t\langle n_a \rangle} + \frac{1}{I_{dd}}.
\end{align}
It shows that the phase sensitivity $\Delta \theta$ in the MZI improves with the average number of photons $\langle n_a \rangle$ when $t$ is close to 1. When PASCS is the input state, the number of photons in the MZI increases due to the addition of the photon(s), which leads to the robust nature (against photon loss) of the MZI working with PASCSs as input states.

\section{Conclusion}\label{sec:Conclusion}
In this paper, we have studied the phase sensitivity in MZI via different input states. We found that PASCS state as input of MZI performs better than PSSCS or a combination of PASCS and PSSCS. We have used QFI to show it, and our investigation shows that QFI is maximum for PASCS input state and minimum for PSSCS input state. QFI for the combined PASCS and PSSCS state is between PASCS and PSSCS, respectively. This establishes that increasing the number of added photons in the system gives the lowest bound on the phase sensitivity of MZI. Therefore, for further analysis, we have used PASCS state as input to the MZI.

The PASCS state has three parameters $(\alpha,r, \phi)$ along with the number of added photons, respectively, where $\alpha$ is the displacement parameter, $r$ is the squeezing amplitude, and $\phi$ is the phase associated with squeezing. $\Delta \theta$ in the MZI with PASCS input states reduces with the displacement parameter $\alpha$ and squeezing amplitude $r$, while it approaches minimum when the squeezing phase is around $\pi$. We have also studied the MZI using PASCS input with IDD and SID detection scheme, to find the conditions where phase sensitivity reaches QCRB obtained via QFI. Due to the uncertainty relation between the intensity and phase, in both IDD and SID, which are mainly intensity detection schemes, phase sensitivity $\Delta \theta $ increases with the intensity in the system. Our analysis is consistent with this fact, and we have observed that with squeezing amplitude $r$, which contributes to the increase in the precise intensity of the MZI, the $\Delta \theta$ in both the detection schemes increases. %The sensitivity in the MZI also increases with the added photons.

We have also studied the use of MZI for phase estimation in the presence of loss, and we found that when the photon loss is small, the QFI in the system is robust against the loss therefore maintain the lower bound on the phase sensitivity. Due to the robustness of the PASCS state against photon loss in MZI, it is interesting to explore this theoretical model for the practical proposes such as its use in quantum radars where the study of attenuation constant with distance is an important parameter. In this work, we have also found that intensity measure is not an optimal detection scheme for phase estimation therefore, it also of great importance to find the optimal detection technique for phase estimation when the input states are PASCSs. Another important extension of this work is to use this PASCS-state in an enhanced MZI-setup such as in an adaptive or feedback scheme to explore the states importance in sensory and amplifier-based applications.

\section*{Acknowledgments}

Authors acknowledge the support from the QUEST scheme of Interdisciplinary
Cyber Physical Systems (ICPS) program of the Department of Science
and Technology (DST), India (Grant No.: DST/ICPS/QuST/Theme-1/2019/14
(Q80)).

\bibliography{MZI}

%-----------------------------APPENDIX--------------------------------------------
%\newpage

\end{document}